# Modeling Space-Charge Limited Currents in Organic Semiconductors: Extracting Trap Density and Mobility

J. Dacuña[1], A. Salleo[2]

[1]Department of Electrical Engineering, Stanford University, Stanford CA 94305

[2]Department of Materials Science and Engineering, Stanford University, Stanford CA 94305

## Abstract

We have developed and applied a mobility edge model that takes into account drift and diffusion currents to characterize the space charge limited current in organic semiconductors. The numerical solution of the drift-diffusion equation allows the utilization of asymmetric contacts to describe the built-in potential within the device. The model has been applied to extract information of the distribution of traps from experimental current-voltage measurements of a rubrene single crystal from Krellner et al. [Phys. Rev. B, 75(24), 245115] showing excellent agreement across several orders of magnitude of current. Although the two contacts are made of the same metal, an energy offset of 580 meV between them, ascribed to differences in the deposition techniques (lamination vs. evaporation) was essential to correctly interpret the shape of the current-voltage characteristics at low voltage. A band mobility 0.13 cm$^2$/Vs for holes was estimated, which is consistent with transport along the long axis of the orthorhombic unit cell. The total density of traps deeper than 0.1 eV was $2.2 \times 10^{16}$ cm$^{-3}$. The sensitivity analysis and error estimation in the obtained parameters shows that it is not possible to accurately resolve the shape of the trap distribution for energies deeper than 0.3 eV or shallower than 0.1 eV above the valence band edge. The total number of traps deeper than 0.3 eV however can be estimated. Contact asymmetry and the diffusion component of the current play an important role in the description of the device at low bias, and are required to obtain reliable information about the distribution of deep traps.

# I. Introduction

Organic semiconductors are the object of intense investigation because of the promise of low-cost and large-area electronic applications such as radio-frequency identification (RFID) tags or active matrix display backplanes.[1-7] Large efforts of the research community have led to a constant improvement of the carrier mobility over the years. Mobilities as high as 1.4 cm$^2$/Vs for polymers,[8] 3.4 cm$^2$/Vs for polycrystalline films,[9] and up to 30 cm$^2$/Vs for single-crystal-like films of small molecule organic semiconductors,[10] have been achieved in thin film transistors, exceeding that of hydrogenated amorphous silicon and approaching that of polycrystalline silicon.

A major bottleneck towards the design and development of new materials is our lack of fundamental understanding of what limits charge transport in organic semiconductors. In films of crystalline organic semiconductors, the effect of trap states located in the bandgap is commonly observed.[11-16] The energetic distribution of these localized gap states created by disorder, chemical or morphological defects, or impurities affects the performance of organic devices made with these materials. The accurate characterization of this trap distribution and its correlation to the nature of defects is crucial to fully understand the fundamental limits of these materials. In this regard, investigations of highly purified organic single crystals are fundamental, as these constitute model systems where the effect of microstructural features such as grain-boundaries is suppressed. Single-crystal studies might allow us to predict the upper performance bounds of organic semiconductors as well as help design materials with ever increasing carrier mobility.

The observation of traps in organic single crystals is well documented;[17-21] these may be caused by residual impurities, lattice disorder, or defects such as dislocations. Because in high-quality crystals trap densities are extremely low, obtaining direct evidence of their existence is challenging. Indirect evidence can be obtained by measuring their effect on the I-V characteristics of semiconductor devices where the Fermi-level in the semiconductor is made to sweep through trap distributions. Experimental techniques to estimate the trap distribution include the direct fitting of space charge limited current (SCLC) characteristics to transport models and assuming a particular energetic distribution of traps, where Gaussian and exponential distributions are common choices.[11] More sophisticated methods that take advantage of the spectroscopic character of temperature dependent measurements have also been used to obtain the trap distribution without the assumption of any particular shape.[22,23]

We apply a mobility edge (ME) transport model[24] and solve the drift-diffusion equations to numerically extract the energetic distribution of traps near the valence-band edge in a rubrene single crystal using SCLC measurements obtained from literature. This transport model is widely believed to correctly describe the charge transport properties of single crystals and polycrystalline organic thin films.[12,21,25-27] Compared to field-effect devices, the application of this transport model to the analysis of the SCLC data introduces new challenges. Indeed, in SCLC measurements the Fermi-level does not approach the band edges, therefore traps deeper than those probed in transistors affect the diode currents. Since in high-quality crystals deep traps have low concentrations, a major concern in using SCLC measurements to characterize such traps is the understanding of the accuracy and the sensitivity of the measurements. Furthermore, the charge density in the semiconductor is not conveniently controlled as in a transistor by using the gate terminal. Finally, in the absence of the assistance of the gate field in injecting charge, contact effects may dominate the SCLC data.

To properly account for the experimental data, we show that asymmetric contacts must be assumed even if they are nominally made of the same metal. Additionally, we show that the diffusion term of the current, which is usually neglected in analytical models, can be very important in the characterization of the current at low bias and up to a few volts. Finally, we perform a sensitivity analysis and error estimation to determine the confidence with which electronic structure information can be extracted from such measurements.

## II. Drift-diffusion SCLC model

### A. Justification of the drift-diffusion model

Temperature dependent SCLC measurements have been used to characterize the distribution of traps in organic semiconductors.[23,28,29] The most typical configuration for measuring the SCL current in an organic semiconductor consists of applying a potential difference between two parallel plate contacts at both sides of a relatively thick semiconductor layer.

In order to have monopolar SCL current, the devices are prepared such that the injecting contact is ohmic, i.e. the current flow is not limited by injection from the contact but by the space charge formed within the semiconductor layer, and the extracting contact blocks the injection of the opposite-polarity carrier. In this case, if no active traps are present in the semiconductor, the current follows the well-known Mott-Gurney equation

$$J = \frac{9}{8}\mu\varepsilon_0\varepsilon_r \frac{V^2}{L^3}, \quad (1)$$

where $\mu$ is the charge carrier mobility, $\varepsilon_0$ is the free-space permittivity, $\varepsilon_r$ is the dielectric constant of the semiconductor, $V$ is the applied voltage, and $L$ is the separation between the contacts, or thickness of the semiconductor layer.

Variations to this equation have been developed to include the effect of traps with a pre-defined energetic distribution. A common assumption is an exponential trap distribution of the form

$$\rho_{traps}(E) = \frac{N_t}{E_B}\exp\left(\frac{-E}{E_B}\right), \quad (2)$$

where $E_B$ and $N_t$ are the characteristic energy and total trap density, respectively. In this case, the SCL current is approximated by[30]

$$J = q\mu_0 N_H \left(\frac{\varepsilon}{qN_t}\frac{m}{m+1}\right)^m \left(\frac{2m+1}{m+1}\right)^{m+1}\frac{V^{m+1}}{L^{2m+1}}, \quad (3)$$

where $\varepsilon = \varepsilon_0\varepsilon_r$, $m=E_B/kT$, $q$ is the electron charge, $\mu_0$ is the free-carrier mobility, and $N_H$ is the effective density of states in the band.

A general simplification used in the derivation of analytical models is to neglect the diffusion current as long as the applied voltage is larger than few $kT/q$. We will show later that this is not necessarily true.

Even when the contacts in organic hole-only or electron-only devices are made from the same material, the fabrication process of the bottom and top contacts is intrinsically different. The difference between the bottom and top contacts usually determines which of the two will be used to inject or extract carriers. The energetics of metal-organic interfaces have been extensively studied.[31] One of the fundamental differences between organic-on-metal and metal-on-organic interfaces is that the former is usually formed by depositing the organic semiconductor by evaporation or by spin-coating on a metal that has been exposed to atmosphere and is therefore somewhat contaminated: contaminants include hydrocarbons or residual solvent molecules that may be present due to the processing conditions. This contaminant layer effectively passivates the metal surface, and reduces the interaction between the organic molecules and metal, leading to an energy band alignment that is close to the Schottky-Mott limit.

In contrast, metal-on-organic contacts are usually deposited by metal evaporation in vacuum, where the presence of contaminants is greatly reduced. The stronger interaction between the organic surface and the metal often induces a strong Fermi-level pinning and a large interface dipole, causing the energy band alignment to depart from the vacuum level alignment expected from the electronic structure of the isolated metal and semiconductor.

**Figure 1** represents the energy band diagram of an organic semiconductor with two asymmetric contacts. The right contact aligns close to the HOMO band so that hole injection will occur when properly biased, while the left contact forms a considerable energy barrier to the injection of both electrons and holes. When no voltage is applied, the system is in equilibrium, the Fermi-level is constant throughout the semiconductor and a built-in potential develops in the device. If a small positive voltage, smaller than the onset of SCL current, is applied to the right contact, holes will be injected from the right contact and current will flow through the semiconductor in the direction opposite to the built-in electric field. Therefore, in the low bias region current flow cannot be explained without taking into account the diffusion term. The bias region where the diffusion term is important can extend up to several volts depending on materials and experimental conditions.

## B. Model details

We solve the drift-diffusion equations in the semiconductor according to the mobility edge (ME) model.[24] The ME model divides the density of states into mobile states and trap states. Mobile states, with mobility $\mu_0$, are located above (below) the mobility edge in the case of an n-type (p-type) semiconductor. Quasi-equilibrium is assumed everywhere such that the density of mobile holes (for a p-type semiconductor) can be calculated using Fermi statistics as

$$n_m(x) = \int_{-\infty}^{ME(x)} \rho(E) \frac{1}{1+\exp\left(\frac{E_f(x)-E}{kT}\right)} dE, \qquad (4)$$

where $\rho(E)$ is the density of states and $E_f$ is the quasi-Fermi-level for holes. Similarly, the total hole density $n_t(x)$ is calculated by integrating the same equation from $-\infty$ to $+\infty$.

The current in the device is given by

$$J(x) = q\mu_0 n_m(x) E(x) - qD\frac{dn_m(x)}{dx}, \qquad (5)$$

where $D$ is the diffusion coefficient. Since the current in the device is assumed to be mono-polar, current continuity is given by $dJ(x)/dx = 0$. The charge distribution is then related to the potential variation in the device using Poisson's equation

$$\frac{d^2V}{dx^2} = -\frac{q}{\varepsilon_0 \varepsilon_r} n_t(x). \qquad (6)$$

Finally, the coupled set of differential equations is solved using finite differences. As boundary conditions, we fix the energy difference between the ME and the quasi-Fermi-level position at the contacts.

## III.  Experimental data analysis

### A.  Rubrene crystals current-voltage characteristics

To test the validity of the model, we have used measurements of the SCL current in a rubrene crystal taken from the literature.[23] In this work, the authors grew platelike rubrene crystals by physical vapor transport under an argon stream, with the direction perpendicular to the surface corresponding to the long axis of the orthorhombic unit cell. We chose this particular data set because of the high quality of the data, which span a large temperature interval (110 K to 200 K), with curves that exhibit smooth variations at all temperatures and very low currents measured at low bias. As a result, this data set enables us to test our model over orders of magnitude in current density and allows us to determine the validity of the physical assumptions we have made.

The crystals, with thickness < 2 μm, were placed on glass substrates with 20 nm Au on 5 nm Cr electrodes, thus forming the bottom contacts by electrostatic adhesion. Top contacts were made by evaporating 20 nm of Au while the substrate was slightly cooled to minimize thermal damage. Due to the intrinsic difference in the fabrication of the top and bottom contacts, an asymmetric behavior of the contacts was reported by the authors who observed the superiority of the laminated versus evaporated contact agreeing with the observation that the energetics of organic-on-metal interfaces closely follow the Schottky-Mott limit.[31]

## B. Fitting pre-defined distributions

The density of states in the HOMO band of a two-dimensional rubrene crystal has been calculated using a semi-classical model.[32] Since the dispersion in the direction of the long axis of the unit cell is negligible, we obtain the DOS for a tri-dimensional crystal dividing by the size of the unit cell in this direction. The localization length of the states within the HOMO band showed a dramatic decrease close to the band edge, thus separating mobile states from traps. Consequently, we approximate the HOMO DOS with a uniform energy distribution with value $10^{21}$ cm$^{-3}$eV$^{-1}$.

We approximate the trap density obtained by TD-SCLC spectroscopy[23] using a piecewise exponential function as shown in **Figure 2**. The energy difference between the position of the Fermi-level and the mobility edge at the injecting $\Phi_i$ and extracting $\Phi_B$ contacts is fixed at 50 meV, such that the current in the device is limited by space charge and not by the contact barriers. The results did not depend on the specific value of the energy barrier. The only model parameter, $\mu_0$, is determined by minimizing a weighted sum of the residuals squared (weighted least squares) of a set of temperature dependent I-V curves according to

$$\min_{\mu_0}\left\{\sum_i w_i^2\left(I_i - \hat{I}_i(\mu_0)\right)^2\right\}, \qquad (7)$$

where $I_i$, and $\hat{I}_i$ are the I-V temperature dependent measurements and model predictions, respectively, and $w_i^2$ is a weighting factor proportional to the inverse of the variance of the data $I_i$. The details of the estimation of the variance and weighting factors are found in Appendix A. The fit in **Figure 3** shows that the model correctly replicates the experimental data at high voltage; however in agreement with our previous concerns, the device behavior is not well reproduced at low bias where contacts and deeper traps play a bigger role. The reason is that in a symmetric device in equilibrium, the carriers injected from the contacts generate a potential barrier due to band bending with a maximum at the center of the device known as the virtual cathode. In the low bias regime, the applied voltage not only reduces the energy barrier but also moves its position towards the injecting contact producing linear I-V characteristics.[33] Although the linear region in the symmetric device extends up to a voltage approximately equal to $10kT/q$, the transition between the linear and SCL region described by expressions ( 1 ) or ( 3 ) extends to a substantial higher voltage, as revealed in the model prediction in **Figure 3**. The particular shape of the measured I-V characteristic at low bias is caused by the asymmetry of the contacts in the device, whose accurate modeling requires to account for the diffusion term in equation ( 5 ), and is considered later.

To account for the built-in potential (caused by the contact asymmetry) in analytical models such as those of equations ( 1 ) or ( 3 ), a voltage $V_0$ is usually subtracted from the applied bias.[34] However, $V_0$ is not the difference between the injecting and extracting contact work-functions but is usually smaller due to the strong band bending produced by the accumulation of carriers near the contacts.[35,36] Consequently $V_0$ depends strongly on the density of traps and temperature, which in turn determines the amount of charge accumulated near the contacts.

We now use an exponential distribution of traps in the gap as defined in equation ( 2 ). In addition, the injecting contact $\Phi_i$ is fixed to 50 meV while no assumptions are made about the energy barrier $\Phi_B$ at the extracting contact. As long as the current is not contact limited, the obtained results do not depend on the specific value chosen for the energy barrier at the injecting contact. Values smaller than 50 meV however, are unlikely to occur due to the large electron transfer from the metal to the organic semiconductor as the Fermi-level approaches the band edge.[31] The parameters are now obtained by minimization of

$$\min_{\mu_0, \Phi_B, E_B, N_t} \left\{ \sum_i w_i^2 \left( I_i - \hat{I}_i(\mu_0, \Phi_B, E_B, N_t) \right)^2 \right\}. \quad (8)$$

**Figure 4** compares the experimental data to the result of the numerical model. The obtained density of traps is compared in **Figure 5** to the density of traps obtained using TD-SCLC spectroscopy.[23] The predicted value of the energy barrier at the extracting contact (evaporated contact) is 640 meV, much larger than the 50 meV of the injecting contact and is of the same order as barriers reported in the literature.[31] Because of the asymmetry of the contacts, the model fits the data much better at low bias. In contrast however, the model does not reproduce the correct behavior at high voltage.

According to equation ( 3 ), the slope of the log(I) vs. log(V) curve depends on the characteristic energy of the exponential describing the trap distribution. Hence, the discrepancy between the measurements and the model indicates that a larger $E_B$ is required to fit the data at high voltage. This observation suggests a gradual decrease of the slope of the energetic distribution of traps at lower energies, which can be achieved using a Gaussian function:

$$\rho_{traps}(E) = \frac{N_t}{\sqrt{2\pi\sigma^2}} \exp\left( -\frac{(E - E_c)^2}{2\sigma^2} \right), \quad (9)$$

were $N_t$, $E_c$, and $\sigma$ are the total number of states, the center energy and the standard deviation of the Gaussian distribution, respectively. Using this trap distribution and the corresponding optimization, similar to equation ( 8 ), we obtain $N_t=1.8\times10^{16}$ cm$^{-3}$, $E_c$=153 meV, $\sigma$=50 meV (**Figure 5**), $\mu_0$=0.13 cm$^2$/Vs, and the same energy offset at the contacts as before. The model now correctly reproduces the behavior of the current-voltage characteristics over the entire bias range (**Figure 6**).

### C. Importance of contact asymmetry and diffusion current

**Figure 7** shows the evolution of the potential in the device and the position of its maximum, or virtual cathode, as the applied voltage increases. It is clear that the potential maximum is fixed at the extracting contact for voltages below ~0.3 V, and gradually moves towards the injecting contact as the voltage increases. For voltages above ~2 V, the position of the virtual cathode is relatively constant and the current-voltage characteristic approaches that of equation ( 3 ). **Figure 8** shows the position of the Fermi-level with respect to the band edge near the extracting contact as a function of the applied voltage. At any other point closer to the injecting contact, the Fermi-level is closer to the band edge. It is seen that for an applied voltage of 2 V the Fermi-level is 250 meV away from the band edge at the maximum temperature and closer at lower temperatures, thus all the states deeper than ~0.3 eV from the band edge will be essentially occupied for V>2 V. Since the movement of the virtual cathode, and hence the current in the device below 2 V, cannot be explained without accounting for the diffusion current term in equation ( 5 ) and the contact asymmetry, we conclude that diffusion current and asymmetric contacts are required in order to obtain reliable information of the energetic distribution of states deeper than 0.3 eV.

The particular voltage at which diffusion current can be neglected, as well as the position of the Fermi-level at this voltage, depend on several parameters such as the temperature, the built-in voltage in the device, the trap density or the device length. Hence one cannot find guidelines of general validity as to when the diffusion current can be neglected and only a full device model can provide this information.

## IV.   Establishing Confidence Intervals

When fitting experimental data with a pre-defined density of states, such as the Gaussian distribution defined in equation ( 9 ), the question arises as to whether the obtained distribution really represent the DOS in the semiconductor. In order to shine some light on the problem we have performed an optimization using a piecewise exponential in which the density of states is divided into slices with exponential variation between every two points ($\rho_i$). The density in each slice can be independently estimated with its corresponding confidence interval. The trap distribution is only defined from 0.1 eV to 0.5 eV. Neglecting the states below 0.1 eV and beyond 0.5 eV is justified since states shallower than 0.1 eV do not play an important role in the current-voltage characteristics (limited by the maximum applied voltage) and the model sensitivity to the energetic distribution of deep states is very small, as shown in Appendix B. In addition we expect the trap density to gradually decrease for deeper energies.

The obtained distribution of traps is compared in **Figure 9** to the Gaussian, and the distribution obtained using TD-SCLC spectroscopy.[23] The current-voltage fit is shown in **Figure 10**. The distribution obtained closely matches the Gaussian distribution in the region from 0.1 eV to 0.3 eV and also agrees well with the trap distribution obtained by TD-SCLC spectroscopy. The model however provides more information, such as a quantitative assessment of the contact asymmetry, the mobility in the band, the sensitivity to traps at different energies, and confidence intervals that allow us to assess the reliability on the obtained parameters and trap distribution at different energies. For this particular crystal, we obtain a mobility $\mu_0=(0.13\pm0.04)$ cm$^2$/Vs. Such a low value is consistent with the fact that in this geometry, the mobility is measured in the slowest crystallographic direction. The energetic barrier at the extracting contact is found to be (630±24) meV, essentially the same that was obtained using the Gaussian distribution, thus being relatively insensitive to the details of the density of states. The total trap density is $(2.2\pm0.87)\times10^{16}$ cm$^{-3}$.

The 95% confidence intervals shown in **Figure 9** illustrate that beyond 0.3 eV the accuracy on the obtained trap distribution is small, with uncertainties of several orders of magnitude. Note that it is at E~0.3 eV that the obtained distributions (piecewise exponential and Gaussian) and the distribution obtained using TD-SCLC spectroscopy depart from each other. This observation, together with the fact that states below 0.1 eV do not substantially affect the current-voltage characteristics, explains why a Gaussian provides an excellent fit to the data while not necessarily being an accurate physical representation of the trap distribution over the entire energy range.

The uncertainty in deep state distribution is in part due to the correlation of the obtained $\rho_i$ in this energy range. Such correlation reveals itself in the fact that increasing the trap density at one energy and decreasing that at a neighboring energy (keeping the total number of states constant) has little effect on the current-voltage characteristics. Hence, this correlation of the parameters produces many deep trap distributions that effectively generate the same fit. As a result, the sensitivity of the model to the shape of the deep trap distribution is greatly reduced. Information about the shape of the trap density is simply not present in the measurements and can therefore not be extracted. By integrating the trap energetic distribution beyond 0.3 eV, the uncertainty due to these correlations is mitigated, thus providing an estimate of the total density of deep traps (E>0.3 eV) ranging between $5.5 \times 10^{13}$ cm$^{-3}$ and $2.9 \times 10^{14}$ cm$^{-3}$.

Although there is not enough information to determine the shallow trap distribution from 0 eV to 0.1 eV, it is not unreasonable to assume continuity at the band edge, consistent with the strongly localized states that appear in the band tail due to local and nonlocal electron phonon coupling.[32] We have verified that adding extra states in the region from the band edge to 0.1 eV does not significantly affect the obtained I-V characteristics.

## V. Conclusions

We have applied a mobility edge model and solved drift-diffusion equations to extract information about the trap distribution of a rubrene single-crystal. The model shows an excellent agreement with the experimental measurements over the entire bias range only if an asymmetry in the contacts is introduced. The accurate characterization of the effect of the contact asymmetry requires taking into account the diffusion current term, which is usually neglected in analytical models. Contact asymmetry and current diffusion have proven to be important for the correct characterization of deep traps.

A sensitivity analysis of the model was carried out, showing that the model is not affected by a reasonable distribution of trap states shallower than 0.1 eV. In addition, due to correlation of the parameters characterizing the distribution of deep traps (E>0.3 eV), the uncertainty in this region is high, as seen by the large confidence intervals obtained from the error analysis. The uncertainty in deep states is partially reduced by integrating the distribution to obtain a total density of deep traps. A better characterization of the deep region can be obtained by increasing the resolution of the measurements in the low bias region –using smaller voltage steps–, increasing the device length to reduce the charge density in the device effectively moving the Fermi-level deeper in the band gap – at the expenses of obtaining a lower current however– or using different techniques designed to be sensitive to deep states such as Photothermal Deflection Spectroscopy (PDS)[37] or the Constant Photocurrent Method (CPM).[38]

The use of our complete model will be instrumental in making SCLC measurements as quantitative as possible in both single crystals and thin films. SCLC measurements are attractive because in principle they are the simplest form of electrical characterization of a semiconductor. SCLC measurements are complementary to FET characterization of materials as they probe a different transport direction as well as a different charge density regime, and therefore a different region of the trap distribution.

**Acknowledgments:** This publication was based on work supported by the Center for Advanced Molecular Photovoltaics (Award No KUS-C1-015-21), made by King Abdullah University of Science and Technology (KAUST). JD also gratefully acknowledges funding from "la Caixa" Foundation.

# Appendix A: Error estimation

### 1. Error estimation in original data

We have used SCL current measurements of a rubrene crystal from literature to test the validity of the proposed model. The data has been extracted graphically from a double logarithmic plot in Ref. 23. We assume that the uncertainty in the graphical extraction method is larger than the uncertainty in the original measurements, which we neglect. We also assume for simplicity that the noise in the extracted current follows a Gaussian distribution with zero mean and standard deviation $\sigma$. Due to the nature of the extracting method, in which both independent parameter (voltage) and dependent parameter (current) are estimated graphically, the uncertainty in both quantities is comparable. We treat these uncertainties as if they were only in the dependent variable by combining both fluctuations according to[39]

$$\sigma_{\log(I)} \propto \sqrt{1+\left(\frac{d\log(I)}{d\log(V)}\right)^2}, \qquad (10)$$

where $\sigma_{\log(I)}$ is the standard deviation of the logarithm of the current. Therefore we obtain the following expression for the standard deviation of the current

$$\sigma_I \propto \frac{1}{w} = I\sqrt{1+\left(\frac{d\log(I)}{d\log(V)}\right)^2}, \qquad (11)$$

which is used as weighting factor ($w$) in the weighted least squares estimation.

## 2. Confidence intervals in obtained parameters

Using standard theory of error propagation,[39,40] and linearizing the model around the solution, we estimate the covariance matrix of the obtained parameters as

$$C_\theta = mse \times [J^T J]^{-1}, \tag{12}$$

where $J$ is the Jacobian matrix of the model

$$J = \begin{bmatrix} \dfrac{dI_1}{d\theta_1} & \dfrac{dI_1}{d\theta_2} & \cdots & \dfrac{dI_1}{d\theta_n} \\ \dfrac{dI_2}{d\theta_1} & & & \vdots \\ \vdots & & & \vdots \\ \dfrac{dI_m}{d\theta_1} & \cdots & \cdots & \dfrac{dI_m}{d\theta_n} \end{bmatrix}, \tag{13}$$

and $mse$ the mean squared error calculated as

$$mse = \frac{1}{m-n} \sum_{i=1}^{m} w_i^2 (I_i - \hat{I}_i)^2, \tag{14}$$

with $I$ and $\hat{I}$ being the measured and estimated values of the current, $m$ the number of measurements, $n$ the number of parameters, and the different $\theta_i$ stand for the parameters in the model.

We estimate the 95% confidence intervals for each parameter by using the Student's $t$ distribution with $m-n$ degrees of freedom, and the respective standard deviation obtained as the square root of the diagonal elements of the covariance matrix.

The obtained estimation of the confidence interval is based on the asymptotic normal distribution of the estimated parameters. In some cases, such as small number of samples, when the parameter is close to a boundary, or when the linear approximation is not good enough, its distribution is markedly not normal or skewed and the determination of the confidence interval using the above method may be inaccurate. This is the case for the confidence intervals obtained for parameters defining the trap distribution for energies beyond 0.3 eV, in which the confidence interval calculated with the above method includes negative values and are unphysically large. In these cases, a more robust confidence interval can be constructed using the profile-likelihood function.[41]

The confidence intervals for deep states have been obtained from the profile likelihood function. Assuming Gaussian noise in the data, the probability of observing a current in a small region around $I$ if the actual parameters are $\theta$ is given by $p(I;\theta)\mathrm{d}I$, where the likelihood function $p(I;\theta)$ is given by

$$p(I;\theta) = \prod_{i=1}^{m}\left[\frac{1}{\sigma_i\sqrt{2\pi}}\right]\exp\left[-\frac{1}{2}\sum_{i=1}^{m}\frac{\left(I_i - \hat{I}_i(\theta)\right)^2}{\sigma_i^2}\right], \qquad (15)$$

and the variance of individual measurements $\sigma_i^2$ is estimated as $\sigma_i^2 = mse/w_i^2$. By gradually changing one of the parameters $\theta_o$ while the rest are optimized (i.e. the likelihood function is maximized), we approximate the profile-likelihood function for a single parameter. In order to reduce the computation time, only the strongly correlated parameters (deep states with E>0.3 eV) are considered in the maximization, while the rest are kept fixed at their optimal value.

Then, the confidence interval is given by

$$2\left[l(\hat{\theta}_o) - l(\theta_o^R)\right] = 2\left[l(\hat{\theta}_o) - l(\theta_o^L)\right] = q_1(1-\alpha), \qquad (16)$$

where $l(\theta_o)$ is the natural logarithm of the profile-likelihood function evaluated at $\theta_o$, $\hat{\theta}_o$ is the estimated value of the parameter, $\theta_o^R$ and $\theta_o^L$ are the right and left boundaries of the $(1-\alpha)\times 100\%$ confidence interval, and $q_1(1-\alpha)$ is the $(1-\alpha)$ quantile of the $\chi^2$ distribution with 1 degree of freedom.

## Appendix B: Sensitivity analysis

We start by discretizing the obtained Gaussian distribution with a series of points $\rho_i$ separated $\Delta E$ (12.5 meV) from each other as shown in **Figure 11**. The total density of states around $\rho_i$ is given by $\Sigma_i$, the area under the curve from $E_i - \Delta E$ to $E_i + \Delta E$. We calculate the change in current due to perturbations in the different $\Sigma_i$ as

$$J_\Sigma = \begin{bmatrix} \dfrac{dI_1}{d\Sigma_1} & \dfrac{dI_1}{d\Sigma_2} & \cdots & \dfrac{dI_1}{d\Sigma_n} \\ \dfrac{dI_2}{d\Sigma_1} & \ddots & & \vdots \\ \vdots & & \ddots & \vdots \\ \dfrac{dI_m}{d\Sigma_1} & \cdots & \cdots & \dfrac{dI_m}{d\Sigma_n} \end{bmatrix}. \qquad (17)$$

We define the sensitivity parameters

$$D_i^2 = \sum_{j=1}^{m}\left(\dfrac{dI_j}{d\Sigma_i}\right)^2, \text{ at } E = E_i, \qquad (18)$$

and

$$S_{i,k}^2 = \sum_{j=1}^{m}\left(\dfrac{dI_j}{d\Sigma_i} - \dfrac{dI_j}{d\Sigma_{i+k}}\right)^2, \text{ at } E = \dfrac{E_i + E_{i+k}}{2}, \qquad (19)$$

where $D_i$ represents the sensitivity of the model to changes in the overall amount of states around $E_i$, and $S_{i,k}$ represents the sensitivity of the model to the redistribution of states between $E_i$ and $E_{i+k}$ (i.e. move $\Delta\Sigma$ states from $E_{i+k}$ to $E_i$ such that the total number of states remains constant). Hence $S_{i,k}$ allows us to estimate the sensitivity of the model to the shape of the trap distribution.

**Figure 12** illustrates the variation of the sensitivity of the model, as defined by equations ( 18 ) and ( 19 ) as a function of energy. For energies deeper than 0.3 eV, the model is highly sensitive to changes in the total number of states while the sensitivity to states redistribution (or shape of the density of states) is small, since these changes will only affect the low bias region of the device. For energies shallower than 0.1 eV, the sensitivity to both the states redistribution and total number of states decreases exponentially. In contrast, the energy range from 0.1 eV to 0.3 eV has high sensitivity to any change in the shape or magnitude of the density of states.

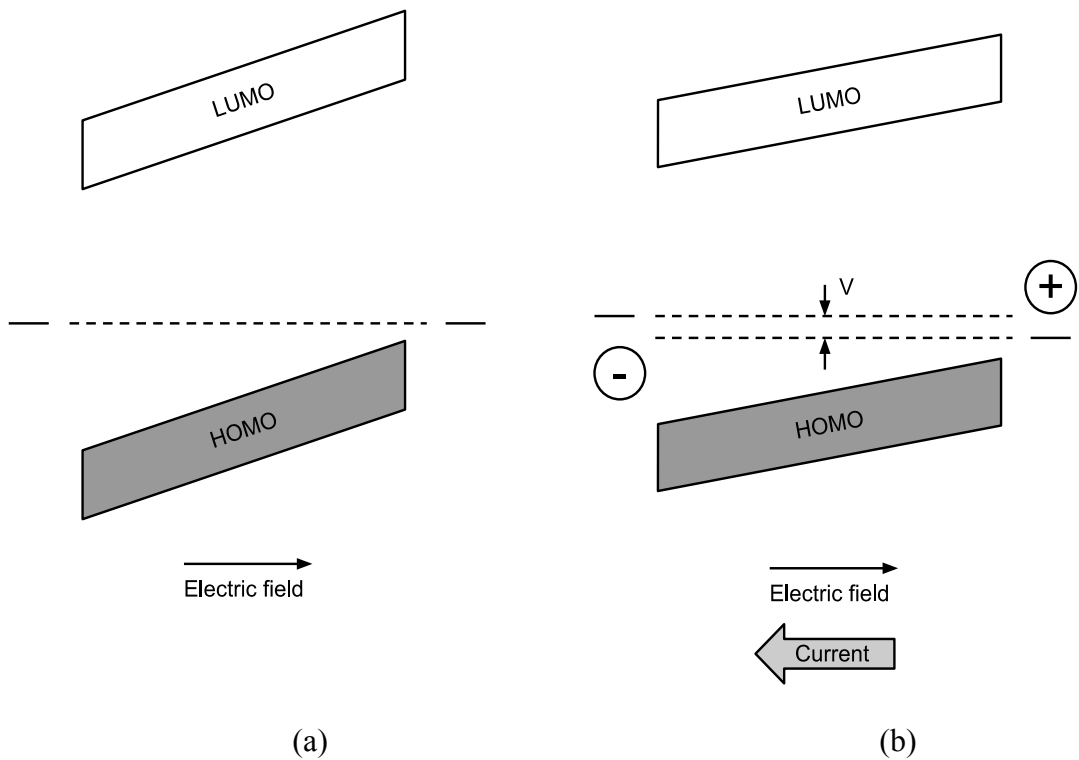

**Figure 1:** Energy band diagram showing the HOMO and LUMO bands of the semiconductor. The horizontal line on both sides represents the position of the Fermi-level at the contacts. (a) System in equilibrium, no current flow. (b) Applying a positive voltage on the right contact make the current flow in the opposite direction of the built-in electric field.

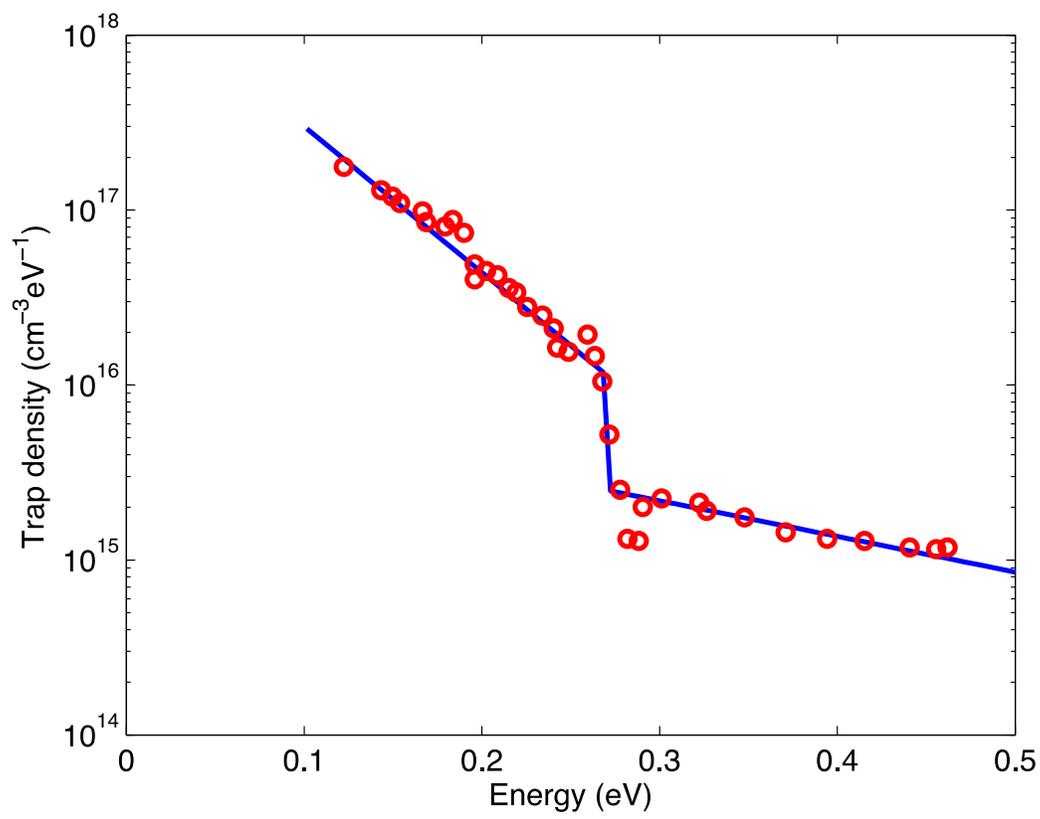

**Figure 2:** Density of trap states calculated in Ref. 23 (red markers) and approximation by a piece-wise exponential (blue line).

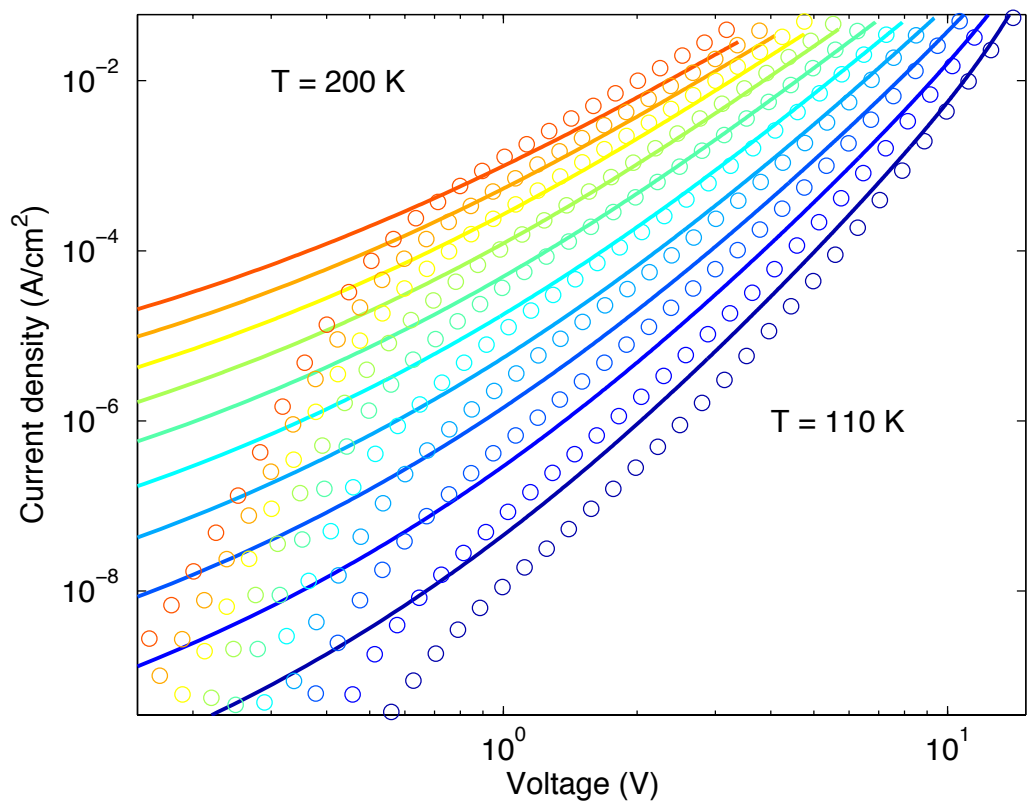

**Figure 3:** Measurement (circles) and model fit using the density of traps in Ref. 23 and symmetric contacts.

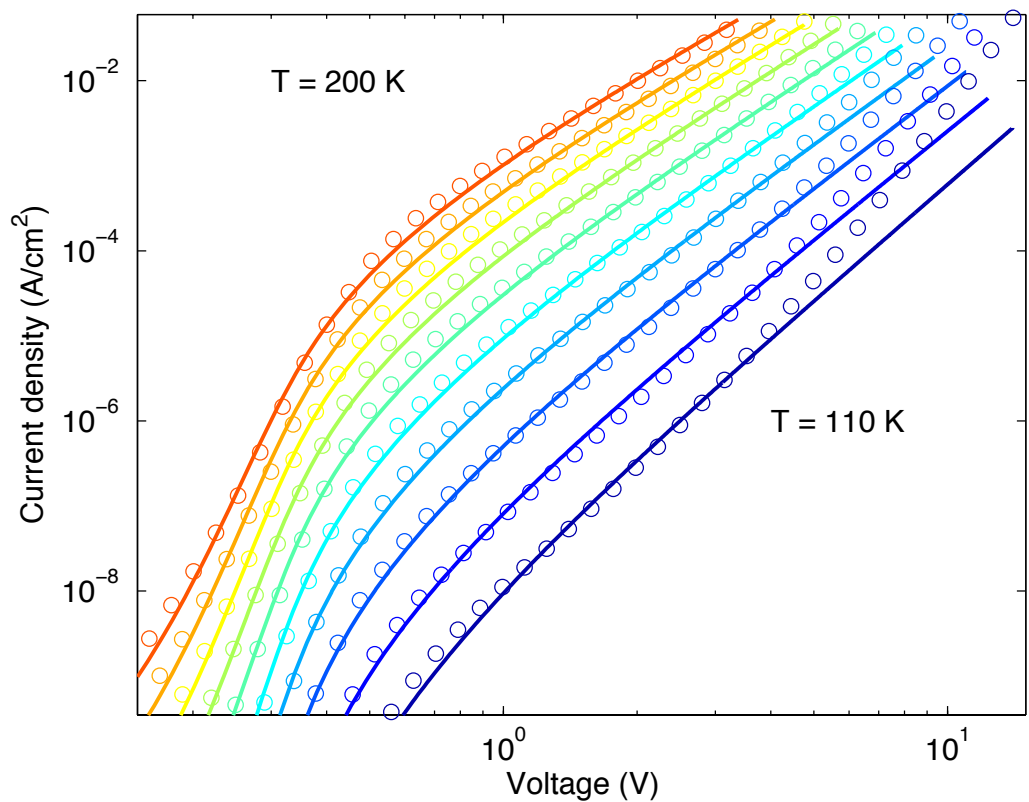

**Figure 4:** Measurement (circles) and model fit using an exponential density of traps and asymmetric contacts.

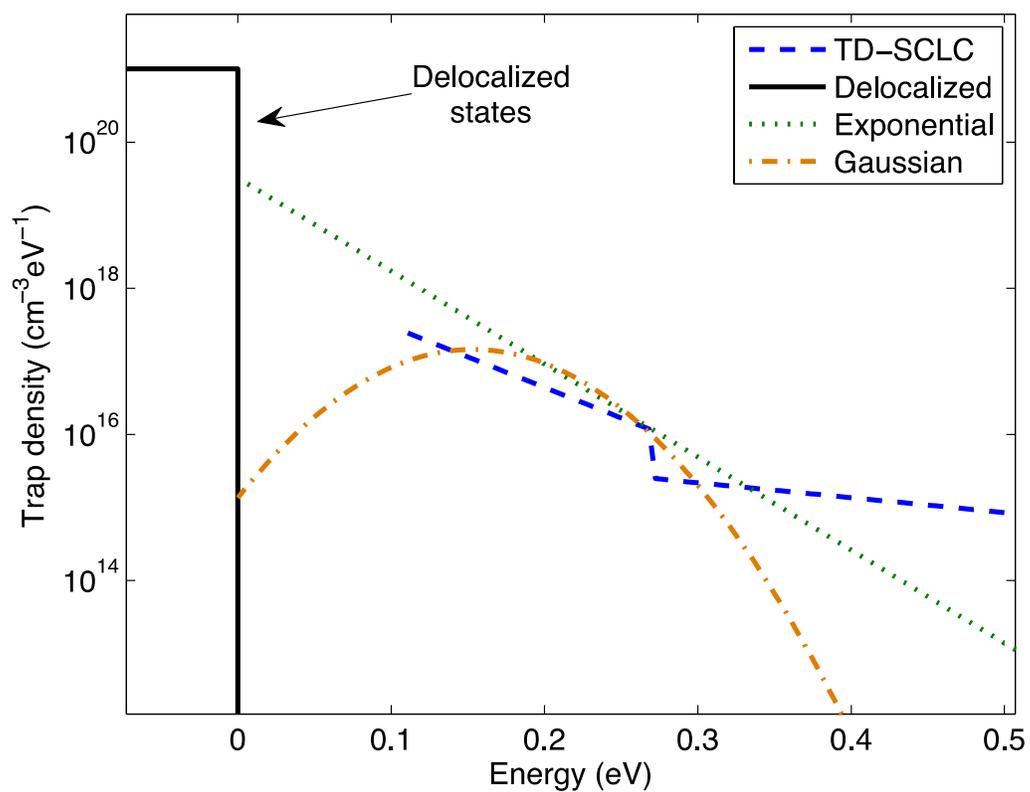

**Figure 5:** Comparison of the obtained exponential and Gaussian density of traps to that obtained in Ref. 23 using TD-SCLC spectroscopy.

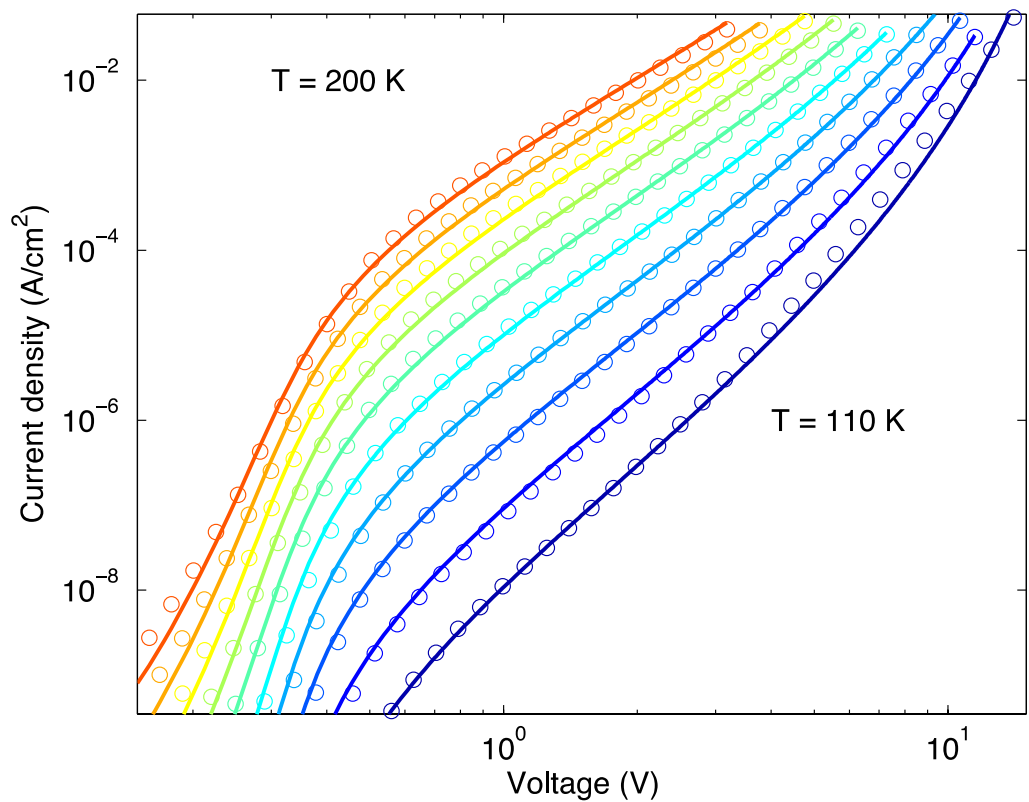

**Figure 6:** Measurement (circles) and model fit using a Gaussian density of traps and asymmetric contacts.

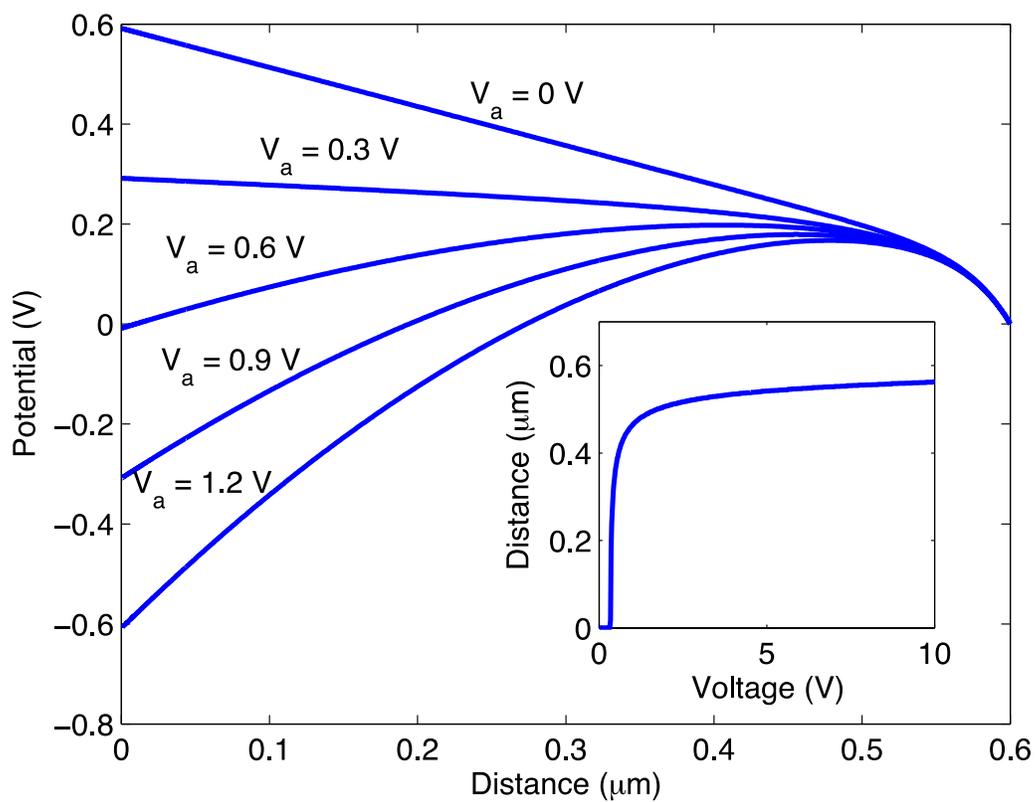

**Figure 7:** Evolution of the potential in the crystal as a function of applied voltage. The inset shows the position of the virtual cathode as a function of the applied voltage.

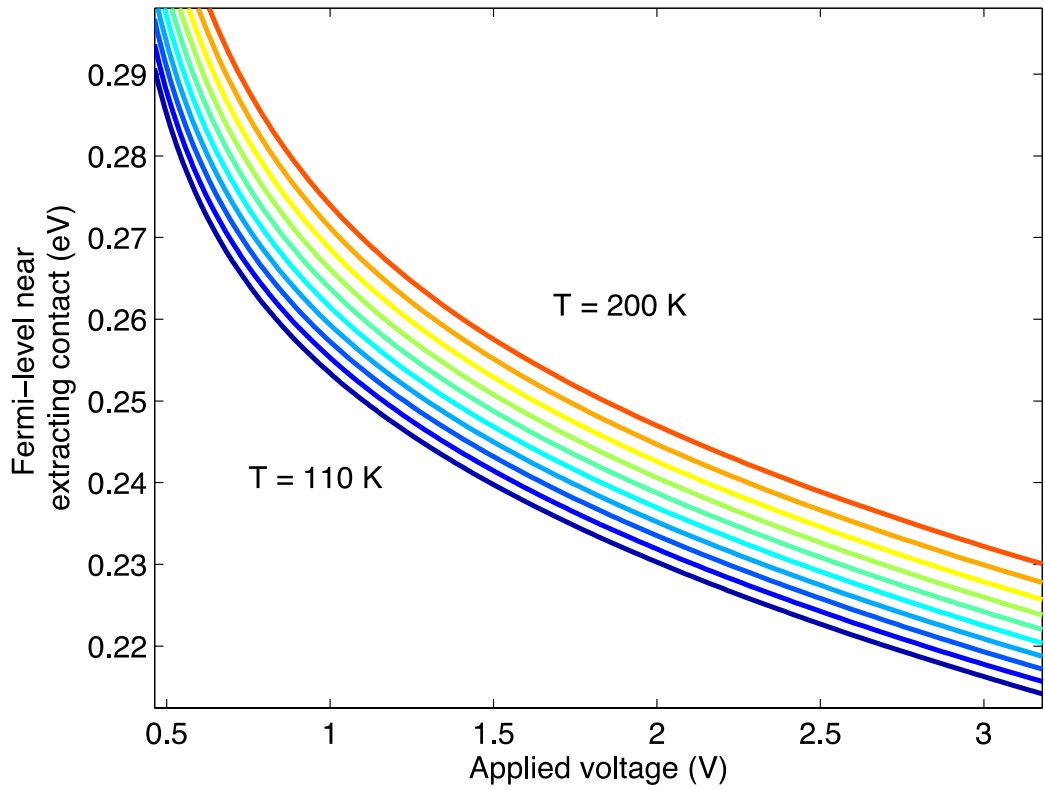

**Figure 8:** Position of the Fermi-level, with respect to the band edge, near the extracting contact as a function of the applied voltage. For voltages beyond 2 V, where the diffusion term can be neglected, the Fermi-level is always below 250 meV, thus all states deeper than ~300 meV are essentially occupied.

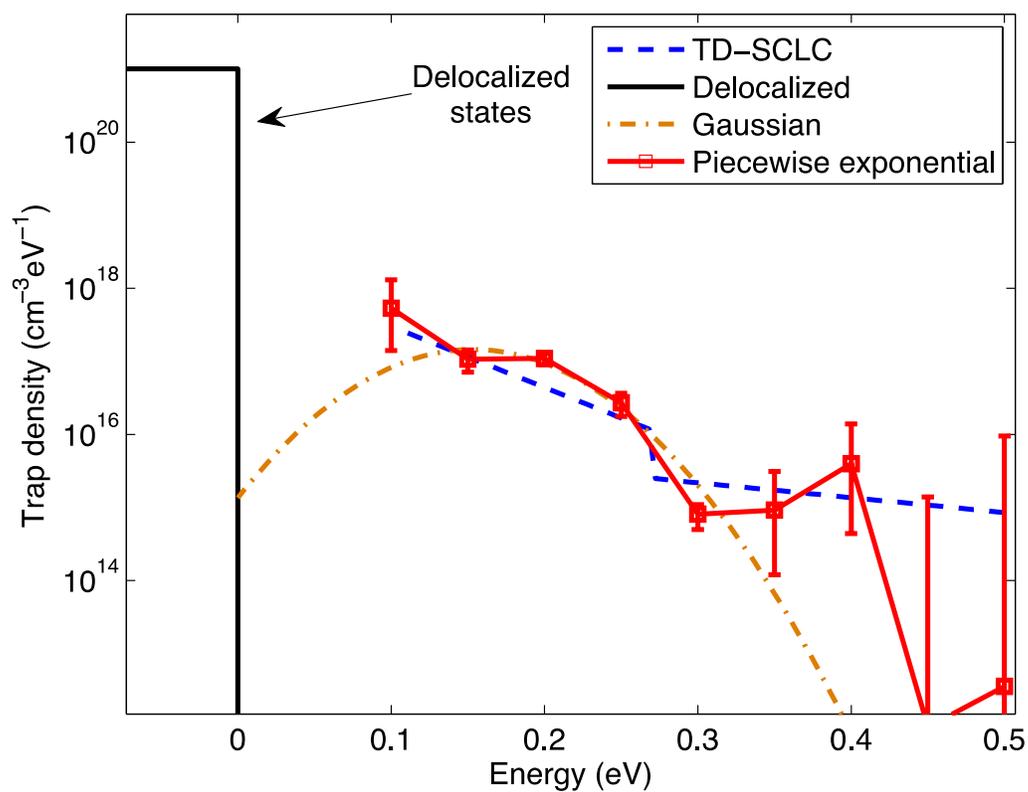

**Figure 9:** Comparison of the obtained Gaussian and piecewise exponential distribution of traps to that obtained in Ref. 23 using TD-SCLC spectroscopy. The three distributions agree in the 0.1 eV to 0.3 eV energy range. Beyond 0.3 eV the distributions diverge and the confidence intervals show that uncertainty in this range increases.

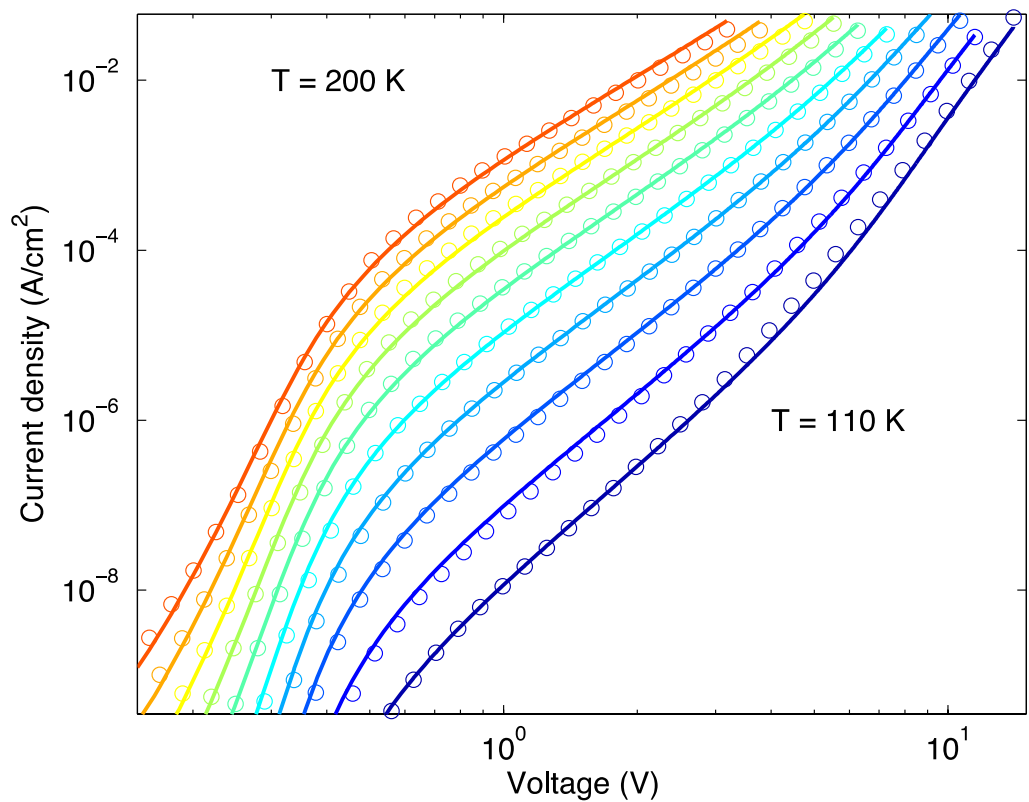

**Figure 10:** Measurement (circles) and model fit using a piecewise exponential density of traps and asymmetric contacts.

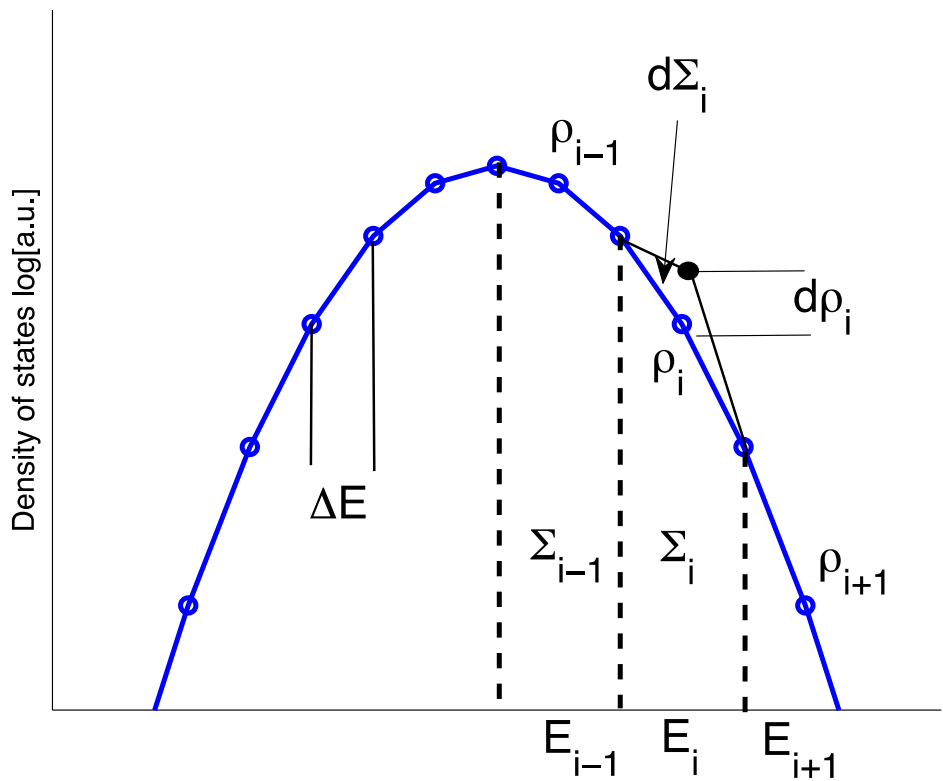

**Figure 11:** Discretization of the Gaussian density of states for the sensitivity analysis.

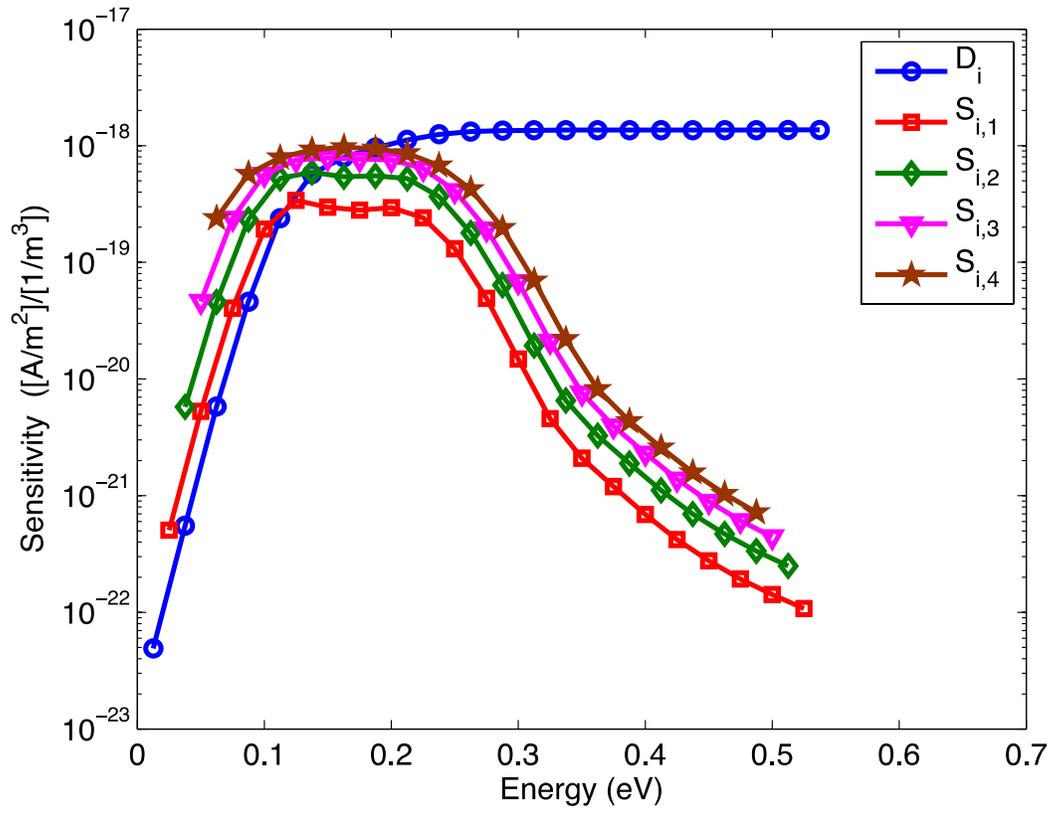

**Figure 12:** Sensitivity of the model to variations in the density of states. D represents the sensitivity to variations in the total density of traps, while S represents the sensitivity to the redistribution of traps (i.e. the shape of the trap distribution) keeping the total number constant.